\documentclass[reprint,superscriptaddress,showpacs,preprintnumbers,amsmath,amssymb,nofootinbib]{revtex4-2}
\usepackage{graphicx}
\usepackage{dcolumn}
\usepackage{bm}
\usepackage{amsthm,amsmath,amssymb}
\usepackage{mathrsfs}
\usepackage{multirow}
\usepackage[T1]{fontenc}
\usepackage{blindtext}
\usepackage[
colorlinks=true,        % color link
citecolor=blue,         % cite color
linkcolor=blue,         % link color
urlcolor=blue           % url color
]{hyperref}             % create hyperlinks
\usepackage{float}

\begin{document}

\newcommand{\be}[1]{\begin{equation}\label{#1}}
	\newcommand{\ee}{\end{equation}}
	\newcommand{\bea}{\begin{eqnarray}}
	\newcommand{\eea}{\end{eqnarray}}
	\newcommand{\bed}{\begin{displaymath}}
	\newcommand{\eed}{\end{displaymath}}
	\def\disp{\displaystyle}
	
% 	\def\gsim{ \lower .75ex \hbox{$\sim$} \llap{\raise .27ex \hbox{$>$}} }
% 	\def\lsim{ \lower .75ex \hbox{$\sim$} \llap{\raise .27ex \hbox{$<$}} }

% \preprint{APS/123-QED}

\title{Towards a reliable reconstruction of the power spectrum of primordial curvature perturbation on small scales from GWTC-3}

\author{Li-Ming Zheng}
\affiliation{Department of Astronomy, Beijing Normal University, Beijing 100875, China}

\author{Zhengxiang Li}
\affiliation{Department of Astronomy, Beijing Normal University, Beijing 100875, China}
\affiliation{Institute for Frontiers in Astronomy and Astrophysics, Beijing Normal University, Beijing
102206, China}

\author{Zu-Cheng Chen}
\affiliation{Department of Astronomy, Beijing Normal University, Beijing 100875, China}
\affiliation{Advanced Institute of Natural Sciences, Beijing Normal University, Zhuhai 519087, China}

\author{Huan Zhou}
\email{zhouh237@mail2.sysu.edu.cn}
\affiliation{School of Physics and Astronomy, Sun Yat-sen University, Zhuhai 519082, China}

\author{Zong-Hong Zhu}
\email{zhuzh@bnu.edu.cn}
\affiliation{Department of Astronomy, Beijing Normal University, Beijing 100875, China}
\affiliation{Institute for Frontiers in Astronomy and Astrophysics, Beijing Normal University, Beijing
102206, China}
%\affiliation{Advanced Institute of Natural Sciences, Beijing Normal University, Zhuhai 519087, China}
%\affiliation{School of Physics and Technology, Wuhan University, Wuhan 430072, China}
\date{\today}

\begin{abstract}
Primordial black holes (PBHs) can be both candidates of dark matter and progenitors of binary black holes (BBHs) detected by the LIGO-Virgo-KAGRA collaboration. Since PBHs could form in the very early Universe through the gravitational collapse of primordial density perturbations, the population of BBHs detected by gravitational waves encodes much information on primordial curvature perturbation. In this work, we take a reliable and systematic approach to reconstruct the power spectrum of the primordial curvature perturbation from GWTC-3, under the hierarchical Bayesian inference framework, by accounting for the measurement uncertainties and selection effects. In addition to just considering the single PBH population model, we also report the results considering the multi-population model, i.e., the mixed PBH and astrophysical black hole binaries model. We find that the maximum amplitude of the reconstructed power spectrum of primordial curvature perturbation can be $\sim2.5\times10^{-2}$ at $\mathcal{O}(10^{5})~\rm Mpc^{-1}$ scales, which is consistent with the PBH formation scenario from inflation at small scales.
\end{abstract}

\maketitle

\section{\label{sec1}Introduction}
The detection of the first gravitational-wave (GW) event GW150914~\citep{Abbott2016} from a binary black hole (BBH) merger opened a new window into astronomy. Up to now, there are $90$ compact binary coalescence candidates reported in the third Gravitational-Wave Transient Catalog (GWTC-3)~\cite{GWTC3} by LIGO-Virgo-KAGRA (LVK) collaboration, of which most are binary black holes. The origin of these black holes is still unknown and under intensive investigation. One of the fascinating possible explanations is the primordial black holes (PBHs) which could form in the early Universe through the gravitational collapse of primordial density perturbations~\cite{Hawking1971,Carr1974,Carr1975}. In order to form PBHs, the amplitude of the power spectrum of primordial curvature perturbations should be larger than $\mathcal{O}(10^{-9})$ measured at $\mathcal{O}(10^{-4}-10^0)~\rm Mpc^{-1}$ scales, e.g., via cosmic microwave background (CMB)~\cite{CMB2018}. If the power spectrum for the primordial curvature perturbations can be enhanced to $\mathcal{O}(10^{-2}-10^{-1})$ at some small scales, it would produce enough PBHs to make up a considerable fraction of dark matter in the Universe~\cite{Sasaki2018, Green2021}. Meanwhile, it also suggests that PBHs would be overproduced when the power spectrum of the curvature perturbation reaches about $10^{-1}$~\citep{Green2021}. Therefore, there are many inflation models, e.g., single-field inflation model with special potentials~\cite{Cai2020,Motohashi2020}, multi-field inflation model~\cite{Clesse2015,Cai2019}, inflation model with modified gravity~\cite{Pi2018,Fu2019}, to enhance the amplitude of power spectrum of primordial curvature perturbations at all kinds of small scales. Theoretically, the mass of PBHs can range from the Planck mass ($10^{-5}~\rm g$) to the level of the supermassive black hole in the center of the galaxy. So far, numerous methods have been proposed to constrain the abundance of PBHs at present in various mass windows. These constraints could be roughly classified into two categories, i.e., direct observational constraints and indirect ones~\cite{Sasaki2018, Green2021}.  

Although the scenarios with stellar mass PBH formation are compatible with the CMB large-scale observations, we still do not know what kind of primordial curvature perturbations could produce enough PBHs to explain current GW events from BBH. Recently, under the expectation that (some of) detected BBH merges could be attributed to PBHs, PBH binaries scenario of currently available GW detection has been proposed to reconstruct the power spectrum of primordial curvature perturbations on small scales~\cite{Kimura2021,Wang2022}. However, these works ignored the selection bias of GW detectors and the individual measurement uncertainty of each GW event. These points are essential for reconstructing the power spectrum of primordial curvature perturbations. In this paper, we proposed to reconstruct the power spectrum of the primordial curvature perturbation in a more reasonable way, which takes individual measurement uncertainty and selection effect for GW observations into account. In addition, we consider two scenarios, i.e., the single PBH population and the multiple-populations model (mixed PBH and astrophysical black hole (ABH) model), under the hierarchical Bayesian inference (HBI) framework to obtain the PBH population hyperparameters. Here, we use the latest 69 GW events of BBH from GWTC-3 to reconstruct the power spectrum combined with the method of HBI. 

This paper is organized as follows. In Section~\ref{sec2}, we introduce the reconstruction method for the  power spectrum of the primordial curvature perturbation. In Section~\ref{sec3}, we present the results of the reconstructed power spectrum. Finally, we present a summary in Section~\ref{sec4}. In this work, we use the concordance $\Lambda$CDM cosmology with the best-fitting parameters from the recent $Planck$ observations~\citep{planck2018}.

\section{\label{sec2}Analysis setup}
In this section, we summarise the population models, i.e., the PBH and ABH binaries model, and statistical framework for the latest GWTC-3 dataset.

\subsection{Population Models}
%PBH
For the PBH binaries model, there are two distinct mechanisms to form PBH binaries theoretically. The first mechanism operates by decoupling from the cosmic expansion in the early Universe dominated by radiation~\cite{Sasaki2016,Haimoud2017,Raidal2017,Chen2018,Raidal2018}. The second one is that PBH binaries form in the late Universe by the close encounter~\cite{Sasaki2016,Raidal2017}. Compared with the second formation mechanism, the mergers from the first channel contribute dominant GW sources of BBH~\cite{Sasaki2016,Raidal2017}. Therefore, for consistency and illustration, here we apply the first formation mechanism of PBH binaries, which usually corresponds to the differential merger rate density as~\cite{Raidal2018}
\begin{equation}\label{eq2-1}
\begin{split}
&\mathcal{R}_{\rm PBH}(\lambda|\theta,f_{\rm PBH})=\frac{1.6\times10^6}{\mathrm{Gpc}^{3}\mathrm{yr}}f_{\rm PBH}^{\frac{53}{37}}\bigg(\frac{t(z)}{t_0}\bigg)^{-\frac{34}{37}}\times\\
&\eta^{-\frac{34}{37}}\bigg(\frac{M}{M_{\odot}}\bigg)^{-\frac{32}{37}}S(M,f_{\rm PBH},P_{\rm PBH}(m|\theta),z)\times\\
&P_{\rm PBH}(m_1|\theta)P_{\rm PBH}(m_2|\theta),
\end{split}
\end{equation}
where $\lambda\equiv[m_1,m_2,z]$ are the parameters measured by LVK, $M=m_1+m_2$, $\eta=m_1m_2/M$, $\theta$ denotes the population hyperparameters, $P_{\rm PBH}(m|\theta)$ is the normalized mass function of the PBH, $f_{\rm PBH}$ is the abundance of PBH in the dark matter, and $t_0$ is the age of the Universe. Here, $S(M,f_{\rm PBH},P_{\rm PBH}(m|\theta),z)<1$ is a suppression factor including two effects, i.e., the effect of the surrounding smooth matter component on the PBH binary formation and the disruption of the PBH binary by other PBH clusters. We can separately define each contribution as
\begin{equation}\label{eq2-1-1}
\begin{split}
S\equiv S_1(M,f_{\rm PBH},P_{\rm PBH}(m|\theta))S_2(f_{\rm PBH},z).
\end{split}
\end{equation}
An analytic expression for suppression factor can be found in Refs.\citep{Raidal2018,Hutsi2021,Luca2021,Franciolini2022b}. The first term $S_1$ in Eq.~(\ref{eq2-1-1}) could be approximate to
\begin{equation}\label{eq2-1-2}
\begin{split}
S_1\approx 1.42\bigg[\frac{\langle m^2\rangle/\langle m\rangle^2}{\bar{N}+C}+\frac{\sigma_{\rm m}^2}{f_{\rm PBH}^2}\bigg]^{-\frac{21}{74}}\exp(-\bar{N}),
\end{split}
\end{equation}
with 
\begin{equation}\label{eq2-1-3}
\begin{split}
\bar{N}\equiv\frac{M}{\langle m\rangle}\bigg(\frac{f_{\rm PBH}}{f_{\rm PBH}+\sigma_{\rm m}}\bigg),
\end{split}
\end{equation}
where $\sigma_{\rm m}\approx 0.004$ is the rescaled variance of matter density perturbations. The constant factor $C$ is defined as 
\begin{equation}\label{eq2-1-4}
\begin{split}
&C\equiv\frac{\langle m^2\rangle f_{\rm PBH}^2}{\langle m\rangle^2\sigma_{\rm m}^2}\times\\
&\bigg\{\bigg[\frac{\Gamma(29/37)}{\sqrt{\pi}}U\bigg(\frac{21}{74},\frac{1}{2},\frac{5f_{\rm PBH}^2}{6\sigma_{\rm m}^2}\bigg)^{-\frac{74}{21}}\bigg]-1\bigg\}^{-1},
\end{split}
\end{equation}
where $\Gamma(x)$ and $U(a,b,z)$ are Gamma function and confluent hypergeometric function respectively. In addition, the mass average $\langle m^n\rangle$ in the Eq.~(\ref{eq2-1-2}) is defined as
\begin{equation}\label{eq2-1-5}
\langle m^n\rangle\equiv \int m^nP_{\rm PBH}(m|\theta)dm.
\end{equation}
Considering the fraction of PBH binaries disrupted by other PBH clusters, we can write the second term $S_2$ in Eq.~(\ref{eq2-1-1}) as
\begin{equation}\label{eq2-1-5}
S_2\approx \min[1,9.6\times10^{-3}x^{-0.65}\exp(0.03\ln^2x)],
\end{equation}
where $x\equiv (t(z)/t_0)^{0.44}f_{\rm PBH}$. This is a good approximation at $z\leq100$. For the model-independent mass distribution of the PBH model in Eq.~(\ref{eq2-1}), some mass functions have the robust physical meaning of the PBH formation mechanism. Here, we take a typical and popular mass function, i.e., the log-normal mass function~\cite{Carr2017, Bellomo2018} in our following analysis
\begin{equation}\label{eq2-2}
P_{\rm PBH}(m|\theta)=\frac{1}{\sqrt{2\pi}\sigma_{\rm c} m}
\exp\bigg(-\frac{\ln^2(m/m_{\rm c})}{2\sigma_{\rm c}^2}\bigg),
\end{equation}
where $m_{\rm c}$ and $\sigma_{\rm c}$ denote the peak mass of $mP_{\rm PBH}(m|\theta)$ and the width of mass spectrum, respectively. Therefore, the hyperparameters are $\theta=[m_{\rm c}, \sigma_{\rm c}]$. This mass function is often a good approximation if the PBHs are produced from a smooth, symmetric peak in the inflationary power spectrum. It has been demonstrated to be viable when the slow-roll approximation holds~\cite{Green2016, Kannike2017}. Therefore the population hyperparameters of PBH model $\Phi_{\rm PBH}$ are
\begin{equation}\label{eq2-3}
\Phi_{\rm PBH}=[m_{\rm c},~\sigma_{\rm c},~f_{\rm PBH}].
\end{equation}

%ABH
Regarding the ABH binaries model, we describe the merger rate as POWER LAW + PEAK~\cite{Talbot2018} adopted by the LVK population analyses~\cite{LVK2022}. We write the differential merger rate density of the ABH model as 
\begin{equation}\label{eq2-4}
\begin{split}
\mathcal{R}_{\rm ABH}(\lambda|R_{\rm 0,ABH},\theta_z,\theta_1,\theta_2)=R_{\rm 0,ABH}\times\\
p^z_{\rm ABH}(z|\theta_z)p^{m_1}_{\rm ABH}(m_1|\theta_1)p^{m_2}_{\rm ABH}(m_2|\theta_2),
\end{split}
\end{equation}
where $R_{\rm 0,ABH}$ is the local ABH merger rate defined as $R_{\rm 0,ABH}\equiv \int dm_1dm_2\mathcal{R}_{\rm ABH}(z=0)$, $\theta_z$ is population hyperparameters for the redshift distribution of ABH binaries, and $\theta_1$ and $\theta_2$ are the population hyperparameters for the mass function of the $m_1$ and $m_2$, respectively. For the redshift model of ABH binaries, a general parameterization of $p^z_{\rm ABH}(z|\kappa,\gamma,z_{\rm p})$ can be written as~\cite{Madau2014}
\begin{equation}\label{eq2-5}
\begin{split}
&p^z_{\rm ABH}(z|\kappa,\gamma,z_{\rm p})=[1+(1+z_{\rm p})^{-\kappa-\gamma}]\times\\
&\frac{(1+z)^{\kappa}}{1+[(1+z)/(1+z_{\rm p})]^{\kappa+\gamma}},
\end{split}
\end{equation}
where $\kappa$ and $\gamma$ describe the low and high redshift power-law slopes, respectively, and $z_{\rm p}$ corresponds to the peak in $p^z_{\rm ABH}(z)$. For GW events in GWTC-3, most of them are detected at low redshift, so $p^z_{\rm ABH}(z|\kappa,\gamma,z_{\rm p})$ can be simplified as~\cite{Fishbach2018}
\begin{equation}\label{eq2-6}
p^z_{\rm ABH}(z|\kappa)=(1+z)^{\kappa}.
\end{equation}
This distribution is adopted in our following analysis. In addition, the distribution of primary binary black hole mass $m_1$ in Eq.~(\ref{eq2-4}) is described by a mixture of a power-law model
\begin{equation}\label{eq2-7}
\begin{split}
&P_{\rm ABH}(m_1|\alpha,m_{\min},m_{\max})\propto m_{1}^{-\alpha}\times\\
&\mathcal{H}(m_1-m_{\min})\mathcal{H}(m_{\max}-m_1),
\end{split}
\end{equation}
and a Gaussian peak
\begin{equation}\label{eq2-8}
\begin{split}
N_{\rm ABH}(m_1|\mu_{\rm m},\sigma_{\rm m})=\frac{1}{\sigma_{\rm m}\sqrt{2\pi}}
\exp\bigg[-\frac{(m_1-\mu_{\rm m})^2}{2\sigma_{\rm m}^2}\bigg].
\end{split}
\end{equation}
The mixing mass distribution of $m_1$ between the two components is dictated by $\lambda_{\rm p}$ as 
\begin{equation}\label{eq2-9}
\begin{split}
&p^{m_1}_{\rm ABH}(m_1|\theta_1)=[
(1-\lambda_{\rm p})P_{\rm ABH}(m_1|\alpha,m_{\min},m_{\max})+\\
&\lambda_{\rm p}N_{\rm ABH}(m_1|\mu_{\rm m},\sigma_{\rm m})]S(m_1|m_{\min},\delta_{\rm m}),
\end{split}
\end{equation}
where the term $S(m|m_{\min},\delta_{\rm m})$ is a smoothing function as
\begin{scriptsize}
\begin{equation}\label{eq2-9f}
S(m|m_{\min},\delta_{\rm m})=\left\{
\begin{aligned}
&0&(m<m_{\min})\\
&[f(m-m_{\min},\delta_{\rm m})+1]^{-1}&(m_{\min}\leq m<m_{\min}'),\\
&1&(m\geq m_{\min}')
\end{aligned}
\right.
\end{equation}
\end{scriptsize}
with $m_{\min}'\equiv m_{\min}+\delta_{\rm m}$, and $f(m',\delta_{\rm m})$ is
\begin{equation}\label{eq2-10f}
f(m',\delta_{\rm m})=\exp\bigg(\frac{\delta_{\rm m}}{m'}+\frac{\delta_{\rm m}}{m'-\delta_{\rm m}}\bigg).
\end{equation}
Therefore, the population hyperparameters $\theta_1$ for the mass function of $m_1$ can be taken as $\theta_1=[\alpha,m_{\min},m_{\max}, \mu_{\rm m}, \sigma_{\rm m}, \lambda_{\rm p}, \delta_{\rm m}]$. For a given $m_1$, the secondary mass follows a truncated power-law between $(m_{\min}, m_1)$ with a slope $\beta$, which also includes the smoothing term $S(m|m_{\min},\delta_{\rm m})$
\begin{equation}\label{eq2-10}
\begin{split}
p^{m_2}_{\rm ABH}(m_2|\beta)\propto m_{2}^{\beta}S(m_2|m_{\min},\delta_{\rm m})\mathcal{H}(m_1-m_2),
\end{split}
\end{equation}
Therefore, the population hyperparameters of ABH model $\Phi_{\rm ABH}$ are
\begin{equation}\label{eq2-11}
\Phi_{\rm ABH}=[\alpha,m_{\min},m_{\max}, \mu_{\rm m}, \sigma_{\rm m}, \lambda_{\rm p},\beta,R_{\rm 0,ABH},\kappa,\delta_{\rm m}].
\end{equation}

\subsection{Hierarchical Bayesian Inference}
%HBI
Comparing with the method reconstructing the mass function from GWTC-3 observations in~\cite{Wang2022}, we take a more reasonable approach, i.e. the HBI~\cite{Mandel2019,Chen2019,Chen2020,Luca2020,Luca2021,Hutsi2021,Wong2021,Wu2021,Ng2022,Chen2022,Franciolini2022,Franciolini2022b, Liu:2022iuf}, to determine the mass distribution function $P(m|\sigma_{\rm c}, m_{\rm c})$ of PBH model and the abundance of PBHs $f_{\rm PBH}$. The HBI method is usually used to extract the parameters of the underlying distribution based on a set of observations with measurement uncertainty and selection effect. To extract the population hyperparameters $\Phi$ from $N_{\rm obs}$ detections of GW events $d = [d_1,...d_{N_{\rm obs}}]$, the likelihood for $N_{\rm obs}$ BBH events can be rewritten as~\cite{Mandel2019,Chen2019,Chen2020,Luca2020,Luca2021,Hutsi2021,Wong2021,Wu2021,Ng2022,Chen2022,Franciolini2022,Franciolini2022b, Liu:2022iuf}
\begin{equation}\label{eq3-1}
\begin{split}
&p(d|\Phi)\propto N(\Phi)^{N_{\rm obs}}e^{-N(\Phi)\xi(\Phi)}\times\\
&\prod_{i}^{N_{\rm obs}}\int d\lambda L(d_i|\lambda)p_{\rm pop}(\lambda|\Phi),
\end{split}
\end{equation}
where the likelihood of one BBH event $L(d_i|\lambda)$ is proportional to the posterior $p(\lambda|d_i)$. $N(\Phi)$ is the total number of events in the model characterized by the set of population parameters $\Phi$ as
\begin{equation}\label{eq3-2}
N(\Phi)=\int d\lambda T_{\rm obs}\mathcal{R}(\lambda|\Phi)\frac{1}{1+z}\frac{dV_{\rm c}}{dz},
\end{equation}
where $dV_{\rm c}/{dz}$ is the differential comoving volume, the factor $1/(1 + z)$ accounts for the cosmological time dilation from the source frame to the detector frame, and $T_{\rm obs}$ is effective observing time of LIGO O1-O3. In addition, $p_{\rm pop}(\lambda|\Phi)$ is the normalized distribution of black hole masses and redshifts in coalescing binaries as
\begin{equation}\label{eq3-4}
p_{\rm pop}(\lambda|\Phi)=\frac{1}{N(\Phi)}
\bigg[T_{\rm obs}\mathcal{R}(\lambda|\Phi)
\frac{1}{1+z}\frac{dV_{\rm c}}{dz}\bigg].
\end{equation}
Meanwhile, $\xi(\Phi)$ is defined as 
\begin{equation}\label{eq3-5}
\xi(\Phi)\equiv\int d\lambda P_{\rm det}(\lambda)p_{\rm pop}(\lambda|\Phi),
\end{equation}
where $P_{\rm det}(\lambda)$ is the detection probability that depends on the source parameters $\lambda$. We use the simulated signals of injections to estimate the detection fraction~\cite{LIGOo3}. In practice, it is approximately calculated by using a Monte Carlo integral over found injections~\cite{GWTC3}
\begin{equation}\label{eq3-6}
\xi(\Phi)\approx\frac{1}{N_{\rm inj}}\sum_{k=1}^{N_{\rm det}}\frac{p_{\rm pop}(\lambda_k|\Phi)}{p_{\rm draw}(\lambda_k)},
\end{equation}
where $N_{\rm inj}$ is the total number of injections, $N_{\rm det}$ is the number of injections that are successfully detected, and $p_{\rm draw}$ is the probability distribution from which the injections are drawn. Then the posterior distribution $p(\Phi|d)$ can be calculated by
\begin{equation}\label{eq3-7}
p(\Phi|d)=\frac{p(d|\Phi)p(\Phi)}{Z_{\mathcal{M}}},
\end{equation}
where $p(\Phi)$ is prior distribution for the population hyperparameters $\Phi$, and we set prior distributions for all the population hyperparameters $\Phi$ as Table~\ref{tab1}. In addition, $Z_{\mathcal{M}}$ is the Bayesian evidence for the population model $\mathcal{M}$, which can be computed as the integral of the numerator of Eq.~(\ref{eq3-7}) over $\Phi$, i.e.
\begin{equation}\label{eq3-8}
Z_{\mathcal{M}}=\int d\Phi p(d|\Phi)p(\Phi).
\end{equation}
In order to avoid contamination from neutron stars in the GWTC-3~\cite{GWTC3}, we select the BBH merging events satisfying the following criteria: black hole masses ($m_1$ and $m_2$) larger than $3~M_{\odot}$, and inverse false alarm rate ($\rm ifar$) higher than 1 year. Totally there are 69 events from GWTC-3 that satisfy these criteria. Finally, we incorporate the single-population (PBH population) and multi-population (PBH+ABH population) into the \texttt{ICAROGW} \cite{icarogw} to estimate the likelihood function and use \texttt{Bilby} ~\cite{bilby} to search over the parameter space.

In order to compare different models, one can compute the so-called Bayes factor defined as
\begin{equation}\label{eq3-9}
\mathcal{B}^{\mathcal{M}_1}_{\mathcal{M}_2}=\frac{Z_{\mathcal{M}_1}}{Z_{\mathcal{M}_2}}.
\end{equation}
According to Jeffreys scale criterion~\cite{Jeffreys}, a Bayes factor larger than $(10,10^{1.5},10^2)$ would imply strong, very strong, or decisive evidence in favour of model $\mathcal{M}_1$. In addition, we also perform model comparison statistics by using the Bayesian Information Criterion (BIC)~\cite{Schwarz1978} and the Akaike Information Criterion (AIC)~\cite{Akaike1974}. The expressions of the two information criteria are respectively given by
\begin{equation}\label{eq3-10}
\begin{split}
{\rm BIC}_{\mathcal{M}}=-2\ln(\mathcal{L}_{\max})+k\ln(N_{\rm obs}),\\
{\rm AIC}_{\mathcal{M}}=-2\ln(\mathcal{L}_{\max})+2k,~~~~~~~~~\\
\end{split}
\end{equation}
where $k$ represent the total number of population hyperparameters, and  $\mathcal{L}_{\max}$ is the maximum likelihood $p(d|\Phi)$ value for the $N_{\rm obs}$ BBH events. As can be clearly seen from Eq.~(\ref{eq3-10}), population models that give a good fit with fewer parameters will be more favored by GW observations.

\begin{table}[!ht]
\centering
\setlength{\tabcolsep}{5mm}{\begin{tabular}{c|c|c}
\hline
Model & Hyperarameter $\Phi$ & Prior\\
\hline
\multirow{2}{*}{} & $m_{\rm c}$ & $\mathcal{U}[1,~50]$\\ \cline{2-3}
\multirow{2}{*}{PBH} & $\sigma_{\rm c}$  & $\mathcal{U}[0.1,~2]$\\ \cline{2-3}
\multirow{2}{*}{} & $f_{\rm PBH}$ & log-$\mathcal{U}[-5,0]$\\
\hline
\multirow{2}{*}{} & $\alpha$  & $\mathcal{U}[-4,~12]$\\ \cline{2-3}
\multirow{2}{*}{} & $\beta$  & $\mathcal{U}[-4,~12]$\\ \cline{2-3}
\multirow{2}{*}{} & $m_{\min}$  & $\mathcal{U}[2,~10]$\\ \cline{2-3}
\multirow{2}{*}{} & $m_{\max}$  & $\mathcal{U}[50,~200]$\\ \cline{2-3}
\multirow{2}{*}{ABH} & $\mu_{\rm m}$  & $\mathcal{U}[20,~50]$\\ \cline{2-3}
\multirow{2}{*}{} & $\sigma_{\rm m}$  & $\mathcal{U}[1,~10]$\\ \cline{2-3}
\multirow{2}{*}{} & $\lambda_{\rm p}$  & $\mathcal{U}[0,~1]$\\ \cline{2-3}
\multirow{2}{*}{} & $R_{\rm 0,ABH}$  & $\mathcal{U}[0,~200]$\\ \cline{2-3}
\multirow{2}{*}{} & $\kappa$  & $\mathcal{U}[0,~10]$\\ \cline{2-3}
\multirow{2}{*}{} & $\delta_{\rm m}$  & $\mathcal{U}[0,~10]$\\
\hline
\end{tabular}}
\caption{\label{tab1} Population hyperarameters $\Phi$ and their prior distributions used in the HBI. The local ABH merger rate $R_{\rm 0,ABH}$ are in units of $\rm Gpc^{-3}yr^{-1}$, and $[m_{\rm c}, m_{\min}, m_{\max}, \mu_{\rm m}, \sigma_{\rm m}, \delta_{\rm m}]$ are in units of $M_{\odot}$.}
\end{table}

\section{\label{sec3}Reconstruction of primordial curvature perturbation}
%reconstruction
In this section, we introduce and redefine the reconstruction process from the merger rate of the PBH to the primordial curvature perturbation at small scales following Refs.~\cite{Kimura2021,Wang2022}
\begin{equation}\label{eq4-1}
\begin{split}
&\mathcal{R}_{\rm PBH}(\lambda|\theta, f_{\rm PBH})\rightarrow f_{\rm PBH}P(m|\theta)\rightarrow \\
&\sigma^2(m)\rightarrow \mathcal{P}_{\mathscr{R}}(k),
\end{split}
\end{equation}
where $\sigma^2(m)$ is the variance of the density perturbation smoothed by a comoving length scale $R$, and $\mathcal{P}_{\mathscr{R}}(k)$ is the power spectrum of the primordial curvature perturbation. The viability of this process is based on the following three assumptions:
\begin{itemize}
\item Rare high-$\sigma$ peaks of the primordial curvature perturbation in the radiation-dominated era are the main seeds of PBH formation;

\item The window function takes the top-hat form in $k$-space;

\item The probability distribution of the primordial curvature perturbation is $Gaussian$.
\end{itemize}

%\sigma^2-Pk
After obtaining the posterior distributions of parameters $\theta=[\sigma_{\rm c},m_{\rm c}]$ in mass function and $f_{\rm PBH}$ by using HBI, we can derive the variance $\sigma^2(m)$ based on the assumption (3): the distribution of density contrast $\Delta$ follows Gaussian distribution as
\begin{equation}\label{eq4-2}
P(\Delta)=\frac{1}{\sqrt{2\pi\sigma^2}}e^{-(\Delta^2/(2\sigma^2))}.
\end{equation}
Then $\sigma(m)$ can be obtained by the Press-Schechter approach in the non-critical collapse case
\begin{equation}\label{eq4-3}
\sigma(m)=\frac{\Delta_{\rm th}}{\sqrt{2}}\bigg[{\rm erfc}^{-1}\bigg(\frac{2f_{\rm PBH}\Omega_{\rm DM}}{\sqrt{M_{\rm eq}K^3}}m^{3/2}P(m|\theta)\bigg)\bigg]^{-1},
\end{equation}
where $\rm erfc^{-1}(x)$ is the inverse function of the complementary error function,  $\Delta_{\rm th}\approx0.23$ is the threshold of $\Delta$ for PBH formation, $M_{\rm eq}\approx3.52\times10^{17}~M_{\odot}$ is horizon mass at the matter-radiation equality epoch, and $K$ is the ratio between the mass of PBH and horizon mass. In the non-critical collapse case, the mass of PBH is approximately equal to the fraction of horizon mass $M_{k}$ such that $m=KM_{k}$ with $K\approx0.2$~\cite{Carr1975}, so the mass of PBH can be obtained as
\begin{equation}\label{eq4-4}
m(k)=\bigg(\frac{k_{\rm eq}}{k}\bigg)^2M_{\rm eq}K\bigg(\frac{g_{*,\rm eq}}{g_{*}}\bigg)^{1/3},
\end{equation}
where $g_{*}\approx100$ is the number of relativistic degrees of freedom at early universe, and $g_{*,\rm eq}\approx3$ is the  number of relativistic degrees of freedom at matter radiation equality while $k_{\rm eq}=0.01~\rm Mpc^{-1}$. 

The variance $\sigma^2(m)$ is related to the power spectrum of density contrast $\mathcal{P}_{\Delta}$ smoothed by comving scale $R$ with a top-hat window function as
\begin{equation}\label{eq4-4f}
\sigma^2(R)=\langle \Delta_{R}^2\rangle=
\int_0^{\infty}W^2(kR)\mathcal{P}_{\Delta}(t_R,k)d(\ln k).
\end{equation}
Since the power spectrum of density contrast $\mathcal{P}_{\Delta}$ is related to that of curvature perturbation $\mathcal{P}_{\mathscr{R}}(k)$ during the radiation-dominated epoch such that
\begin{equation}\label{eq4-5f}
\mathcal{P}_{\Delta}(t,k)=\frac{16k^4\mathcal{P}_{\mathscr{R}}(k)}{81a^4H^4}, 
\end{equation}
we finally obtain primordial curvature perturbation as~\cite{Kimura2021,Wang2022}
\begin{equation}\label{eq4-5}
\mathcal{P}_{\mathscr{R}}(k)=\frac{81}{16}\bigg(4\sigma^2+k\frac{d\sigma^2}{dk}\bigg)\bigg|_{R=1/k}.
\end{equation}

\begin{figure}
    \centering
	\includegraphics[width=0.5\textwidth]{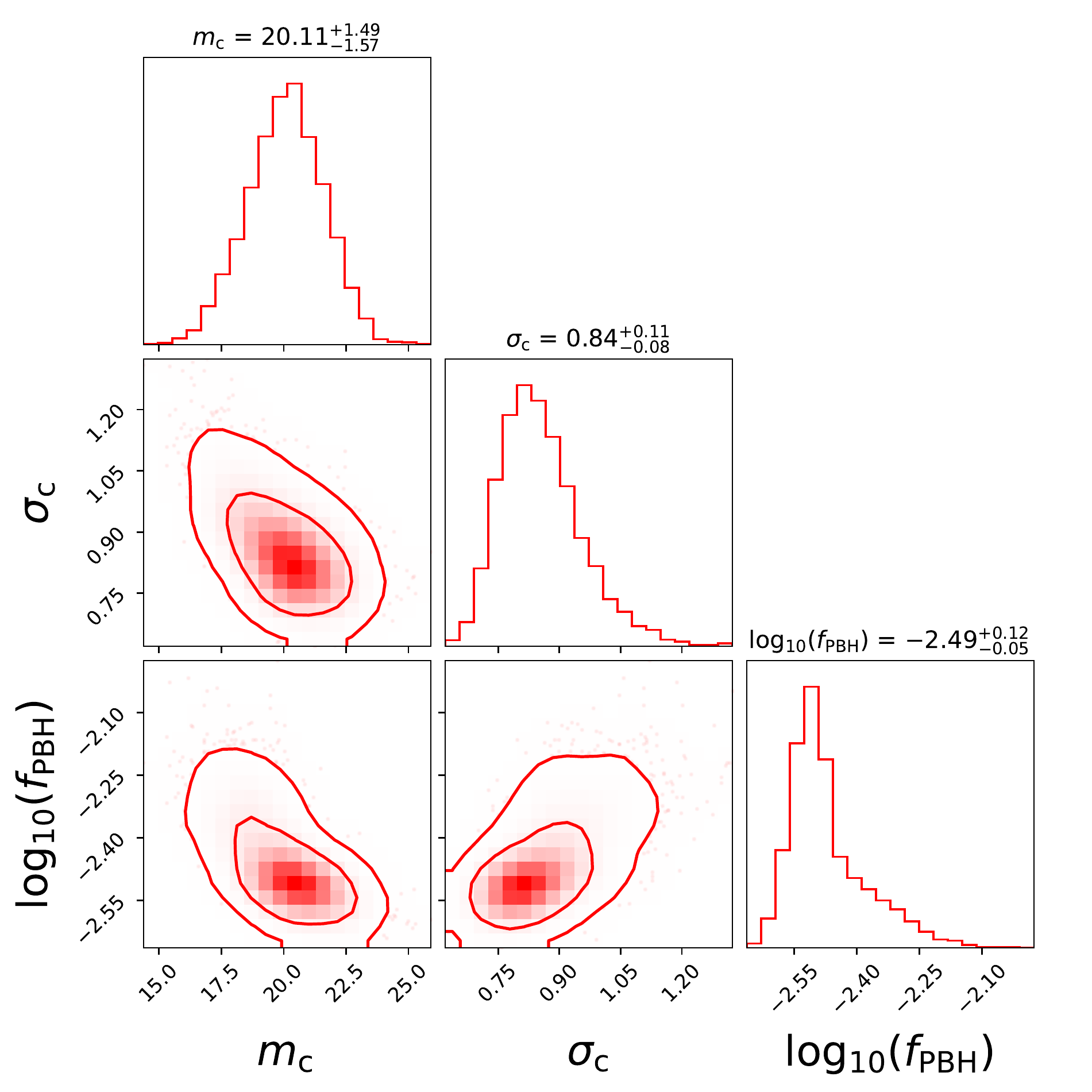}
	\caption{The posterior distributions for $[\sigma_{\rm c}, m_{\rm c}, \log_{10}(f_{\rm PBH})]$ of the log-normal mass function when the single PBH population in the HBI is considered.}
 \label{fig1}
\end{figure}

\begin{figure*}
    \centering
	\includegraphics[width=1\textwidth]{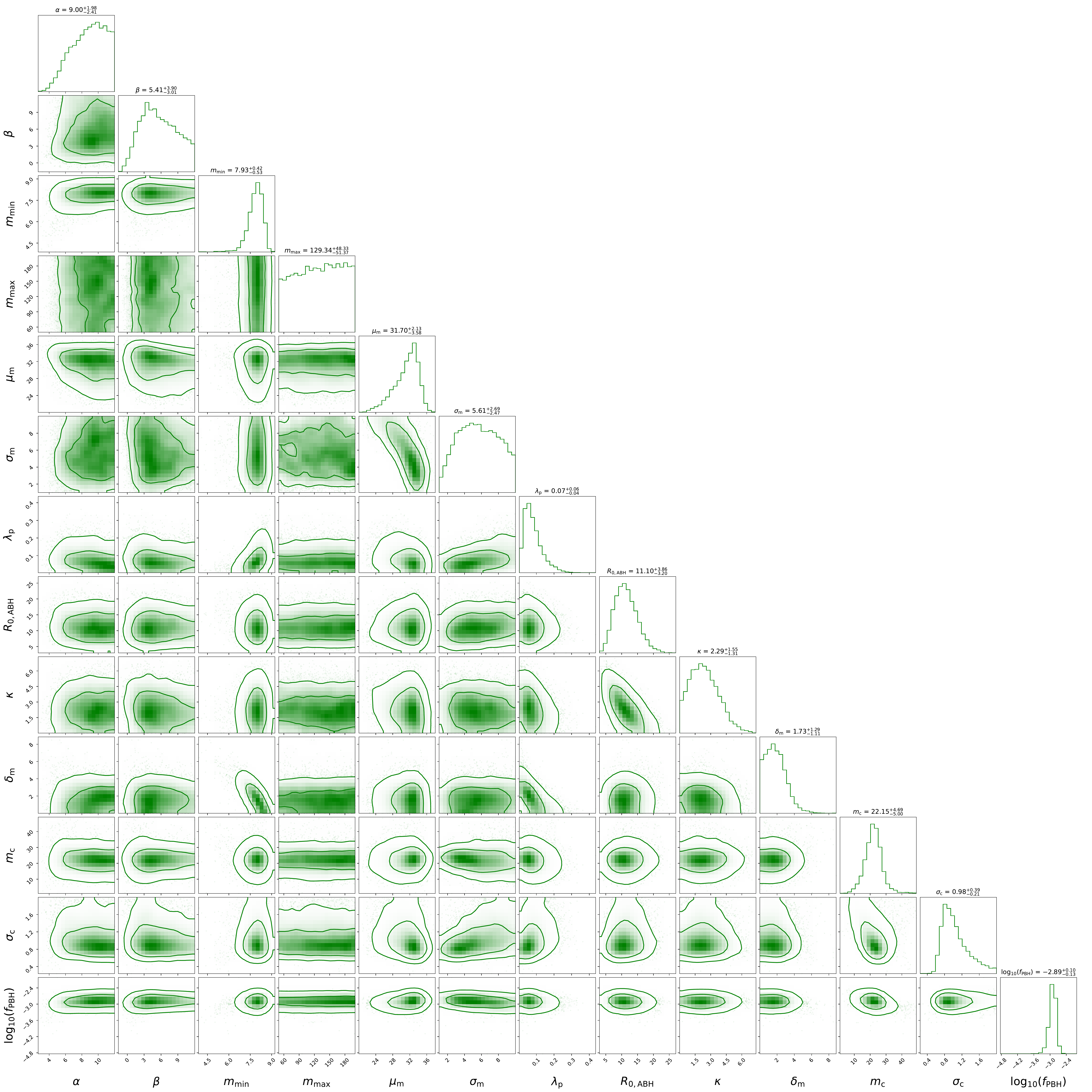}
	\caption{The posterior distributions for population hyperparameters $\Phi=[m_{\rm c},\sigma_{\rm c}, f_{\rm PBH}, R_{\rm 0,ABH},\alpha,m_{\min},m_{\max}, \mu_{\rm m}, \sigma_{\rm m}, \lambda_{\rm p},\beta]$ when the mixed PBH and ABH population model in the HBI is considered.}
 \label{fig3}
\end{figure*}

\begin{figure}
    \centering
	\includegraphics[width=0.47\textwidth]{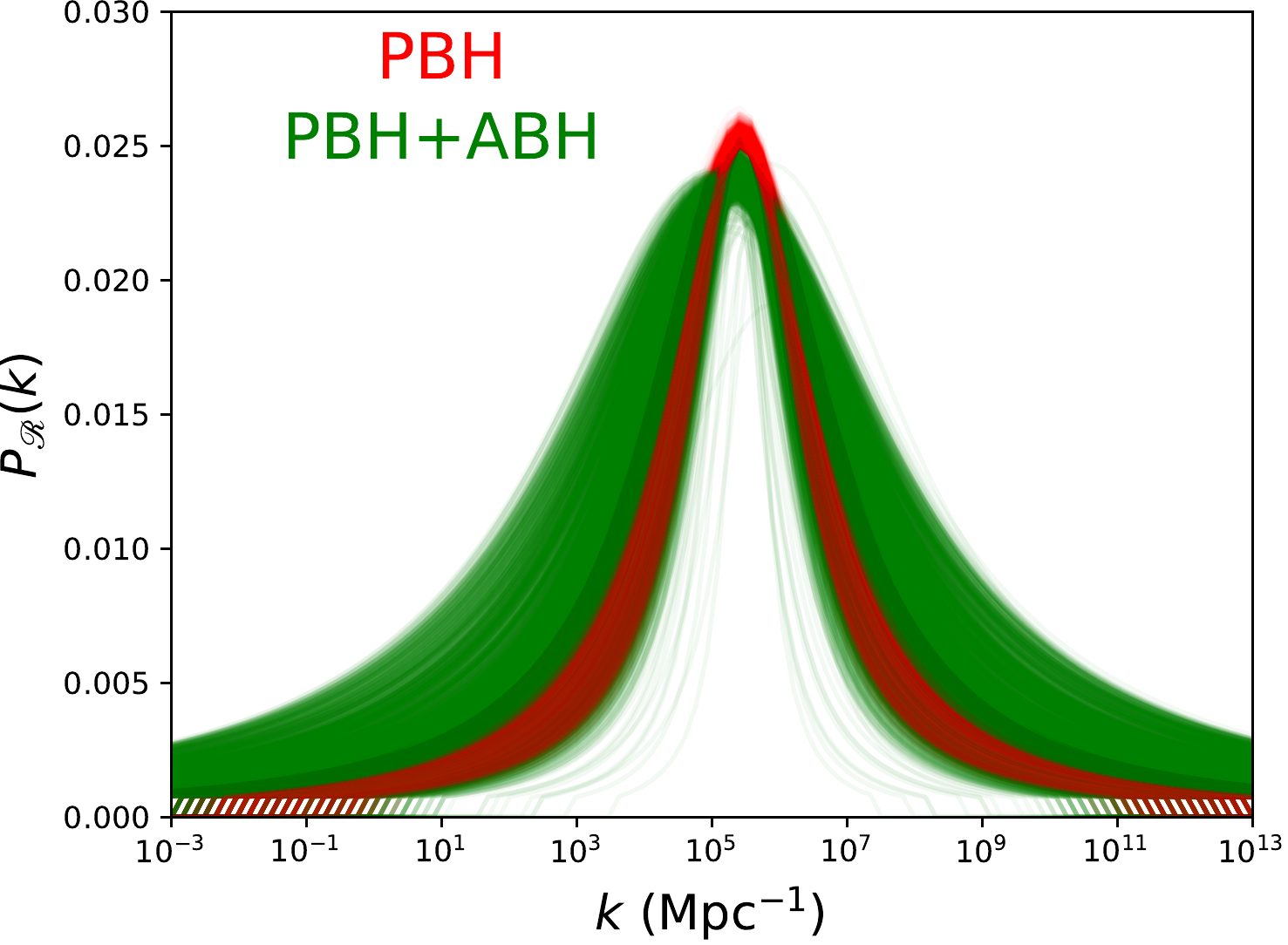}
	\caption{The power spectrum of primordial curvature perturbation derived from the reconstruction procedure with the posterior distributions of $[\sigma_{\rm c}, m_{\rm c}, \log_{10}(f_{\rm PBH})]$. }
 \label{fig2}
\end{figure}

\begin{table*}[!ht]
\centering
\setlength{\tabcolsep}{12mm}{\begin{tabular}{c|c|c}
\hline
Population model & Hyperparameter $\Phi$ & Posterior ($68\%$ C.I.)\\
\hline
\multirow{2}{*}{} & $m_{\rm c}$ & $20.11^{+1.49}_{-1.57}$ \\ \cline{2-3}
\multirow{2}{*}{PBH} & $\sigma_{\rm c}$  & $0.84^{+0.11}_{-0.08}$\\ \cline{2-3}
\multirow{2}{*}{} & $\log_{10}(f_{\rm PBH})$ & $-2.49^{+0.12}_{-0.05}$\\
\hline
\multirow{2}{*}{} & $\alpha$  & $9.00^{+1.98}_{-2.41}$\\ \cline{2-3}
\multirow{2}{*}{} & $\beta$  & $5.41^{+3.90}_{-3.01}$\\ \cline{2-3}
\multirow{2}{*}{} & $m_{\min}$  & $7.93^{+0.42}_{-0.53}$\\ \cline{2-3}
\multirow{2}{*}{} & $m_{\max}$  & $129.34^{+48.33}_{-51.37}$\\ \cline{2-3}
\multirow{2}{*}{} & $\mu_{\rm m}$  & $31.70^{+2.13}_{-3.58}$\\ \cline{2-3}
\multirow{2}{*}{PBH+ABH} & $\sigma_{\rm m}$  & $5.61^{+2.69}_{-2.47}$\\ \cline{2-3}
\multirow{2}{*}{} & $\lambda_{\rm p}$  & $0.07^{+0.06}_{-0.04}$\\ \cline{2-3}
\multirow{2}{*}{} & $R_{\rm 0,ABH}$  & $11.10^{+3.86}_{-3.20}$\\ \cline{2-3}
\multirow{2}{*}{} & $\kappa$  & $2.29^{+1.55}_{-1.31}$\\ \cline{2-3}
\multirow{2}{*}{} & $\delta_{\rm m}$  & $1.73^{+1.29}_{-1.11}$\\ \cline{2-3}
\multirow{2}{*}{} & $m_{\rm c}$ & $22.15^{+4.69}_{-5.00}$ \\ \cline{2-3}
\multirow{2}{*}{} & $\sigma_{\rm c}$  & $0.98^{+0.39}_{-0.21}$\\ \cline{2-3}
\multirow{2}{*}{} & $\log_{10}(f_{\rm PBH})$ & $-2.89^{+0.10}_{-0.13}$\\
\hline
\end{tabular}}
\caption{\label{tab2} Posterior $68\%$ C.I. for popluation hyperparameters $\Phi$ by HBI. `PBH' and `PBH+ABH' represent single-population and multi-population in the HBI, respectively.}
\end{table*}

\section{\label{sec4}Results}
Firstly, we only incorporate the PBH population model $\Phi=\Phi_{\rm PBH}$ and 69 BBH events from GWTC-3 into the \texttt{ICAROGW}~\cite{icarogw} to estimate the posterior Eq.~(\ref{eq3-7}). The posterior distributions of the hyperparameters $\Phi=[m_{\rm c}, \sigma_{\rm c}, \log_{10}(f_{\rm PBH})]$ are shown in Figure~\ref{fig1} and Table~\ref{tab2}. We find the best-fit value and $68\%$ confidence levels for the hyperparameters $[m_{\rm c}, \sigma_{\rm c}, \log_{10}(f_{\rm PBH})]$ to be $m_{\rm c}=20.11^{+1.49}_{-1.57}$, $\sigma_{\rm c}=0.84^{+0.11}_{-0.08}$, $\log_{10}(f_{\rm PBH})=-2.49^{+0.12}_{-0.05}$, which corresponds to the best-fit value of local PBH merger rate $R_{\rm 0,PBH}=\int dm_1dm_2\mathcal{R}_{\rm PBH}(z=0)$ being about $22.6~\rm Gpc^{-3}yr^{-1}$. These results of parameters $[m_{\rm c}, \sigma_{\rm c},f_{\rm PBH}]$ are consistent with the constraints in~\cite{Chen2019,Luca2020,Wu2021,Liu:2022iuf,Wong2021,Chen2022}. In addition, such an abundance $\log_{10}(f_{\rm PBH})=-2.49^{+0.12}_{-0.05}$ of PBHs is consistent with previous estimations that $10^{-3}\leq f_{\rm PBH}\leq10^{-2}$, confirming that most of the dark matter should not consist of stellar mass PBHs~\cite{Sasaki2016,Haimoud2017,Raidal2017,Chen2018}. Then, we use the posterior distributions $[m_{\rm c}, \sigma_{\rm c}, \log_{10}(f_{\rm PBH})]$ shown in Figure~\ref{fig1} to reconstruct the power spectrum of primordial curvature perturbation as Eq~(\ref{eq4-4}). As shown in red lines in Figure~\ref{fig2}, primordial curvature perturbation can be reconstructed at $\mathcal{O}(10^{-3}-10^{13})~\rm Mpc^{-1}$ scales. We find that the maximum amplitude of power spectrum is $2.6\times10^{-2}$ at $\sim3\times 10^{5}~\rm Mpc^{-1}$ scales. 

Moreover, we also derive population inferences assuming multiple channels, i.e. mixed PBH and ABH population models, and 58 BBH events from GWTC-3 and estimate the likelihood function Eq.~(\ref{eq3-7}). The corresponding posterior distribution of hyperparameters $\Phi=\Phi_{\rm PBH}\bigcup\Phi_{\rm ABH}$ are presented in Figure~\ref{fig3} and Table~\ref{tab2}. We obtain the best-fit value and $68\%$ confidence levels for the hyperparameters of PBH $[m_{\rm c}, \sigma_{\rm c}, \log_{10}(f_{\rm PBH})]$ to be $m_{\rm c}=22.15^{+4.69}_{-5.00}$, $\sigma_{\rm c}=0.98^{+0.39}_{-0.21}$, $\log_{10}(f_{\rm PBH})=-2.89^{+0.10}_{-0.13}$. Compared with the best-fit value of the local ABH merger rate $R_{\rm 0,ABH}=11.1~\rm Gpc^{-3}yr^{-1}$, the local PBH merger rate $R_{\rm 0,PBH}$ is about $5.6~\rm Gpc^{-3}yr^{-1}$. In other words, the fractions of detectable events of PBH binaries in the GWTC-3 $f_{\rm p}\equiv N^{\rm det}_{\rm PBH}/(N^{\rm det}_{\rm PBH}+N^{\rm det}_{\rm ABH})$ can be obtained with a peak at $f_{\rm p}\approx 29.7\%$. As shown in Table~\ref{tab3}, we report the Bayes factor comparing the PBH+ABH model to the one which only includes the PBH model; we found Bayes factor is $\log_{10}(\mathcal{B}^{\rm PBH+ABH}_{\rm PBH})=9.88$. According to Jeffreys scale criterion~\cite{Jeffreys}, comparing with the single PBH population model, the Bayes factor shows decisive evidence in favour of the multi-channel population model. These results are consistent with the constraints in~\cite{Luca2021,Hutsi2021,Franciolini2022,Franciolini2022b} by considering multi-population models. For comparing with the single PBH population model, we also show the $\Delta{\rm BIC}$ and $\Delta{\rm AIC}$ in Table~\ref{tab3}. We find that the results of $\Delta{\rm BIC}$ and $\Delta{\rm AIC}$ also show strong favour of the multi-channel population model. Then, we also use the posterior distributions $[m_{\rm c}, \sigma_{\rm c}, \log_{10}(f_{\rm PBH})]$ presented in Figure~\ref{fig3} to reconstruct the power spectrum of primordial curvature perturbation. As shown in green lines in Figure~\ref{fig2}, we find that the maximum amplitude of the power spectrum is from $1.9\times10^{-2}$ to $2.5\times10^{-2}$ at $\sim8\times 10^4-5\times10^5~\rm Mpc^{-1}$ scales.

As shown in Figure~\ref{fig2}, the maximum amplitude of the power spectrum is insensitive to $f_{\rm PBH}$ because the posterior samples range of $f_{\rm PBH}$ does not change significantly. Similarly, the position of the maximum amplitude of the power spectrum is insensitive to $m_{\rm c}$. However, the width of the power spectrum is sensitive to parameter $\sigma_{\rm c}$. The broader power spectrum corresponds to larger $\sigma_{\rm c}$. These results are consistent with the PBH formation scenario that the amplitude of the power spectrum of primordial curvature perturbation is enhanced to $\mathcal{P}_{\mathscr{R}}=\mathcal{O}(10^{-2}-10^{-1})$ at small scales during inflationary epoch. 

\begin{table}[!ht]
\centering
\setlength{\tabcolsep}{3mm}{\begin{tabular}{c|c|c|c|c}
\hline
Population Model & $\mathcal{B}^{\mathcal{M}_1}_{\mathcal{M}_2}$ & $k$ & $\Delta {\rm BIC}$ & $\Delta {\rm AIC}$\\
\hline
PBH & 1 & 3 & 0 & 0\\ \hline
PBH+ABH & $10^{9.88}$ & 13 & -27.62 & -49.96 \\
\hline
\end{tabular}}
\caption{\label{tab3} For comparing the single PBH population model with the multi-population model, we list the Bayes factors, $\Delta{\rm BIC}$, $\Delta{\rm AIC}$. Here, $k$ represents the number of hyperparameters in each population model.}
\end{table}

\section{\label{sec5}Conclusion and discussion}
Based on the scenario that PBH binaries contribute a fraction of the BBH merging events in GWTC-3, we reconstruct the power spectrum of primordial curvature perturbation by using the method of HBI at small scales. We found that the maximum amplitude of power spectrum is $2.5\times10^{-2}$ at $\mathcal{O}(10^{5})~\rm Mpc^{-1}$ scale. Our results are consistent with the theoretical expectation of enhancement of primordial curvature perturbation at small scales for PBH formation. However, there are some uncertainties from several facts. Firstly, the uncertainties of results come from the values for $\Delta_{\rm th}$ which depends on the profile of perturbations, the threshold value of the comoving density contrast could vary from 0.2 to 0.6~\cite{Musco2013,Harada2014,Yoo2018}. This would lead the maximum amplitude of the power spectrum to obtain from $1.4\times10^{-2}$ to $1.7\times10^{-1}$ at $\mathcal{O}(10^5)~\rm Mpc^{-1}$ scales.
Secondly, the effect of the choice of window function would cause the uncertainty in the amplitude of the power spectrum up to $\mathcal{O}(10\%)$~\cite{Gow2020};
Thirdly, our works are based on the Press-Schechter theory and the effects of choice of statistical methods, e.g., Press-Schechter or peaks theory, would slightly affect the results~\cite{Gow2020}; 
Finally, if the primordial curvature perturbation is non-Gaussian, the PBH mass function and PBH abundance would depend on the higher-order statistics~\cite{Yoo2018,Gow2020,Young2013,Pattison2017,Franciolini2018, Atal2019,Luca2022, Kalaja2019,Young2019,Luca2019}. For example, many papers discussing the density contrast $\Delta$ would be non-Gaussian due to the non-linear relationship between the curvature perturbation $\mathscr{R}$ and density contrast $\Delta$~\cite{Yoo2018,Gow2020,Kalaja2019,Young2019, Luca2019}. This effect would lead to the amplitude of the power spectrum of primordial curvature perturbation $\mathcal{P}_{\mathscr{R}}$ must be a factor of $\mathcal{O}(2)$ larger than if we assumed a linear relationship between $\mathscr{R}$ and $\Delta$~\cite{Gow2020,Young2019,Luca2019}.
Therefore, more detailed discussions of the reconstruction procedure are necessary for future work.

\section*{Acknowledgements}
This work was supported by the National SKA Program of China No. 2020SKA0110402; National key R\&D Program of China (Grant No. 2020YFC2201600); National Key Research and Development Program of China Grant No. 2021YFC2203001; National Natural Science Foundation of China under Grants Nos. 12275021, 12021003, 11920101003, 11633001, and 12073088; Guangdong Major Project of Basic and Applied Basic Research (Grant No. 2019B030302001), the Strategic Priority Research Program of the Chinese Academy of Sciences, Grant Nos. XDB2300000 and the Interdiscipline Research Funds of Beijing Normal University.
ZCC is supported by the National Natural Science Foundation of China (Grant No.~12247176) and the China Postdoctoral Science Foundation Fellowship No.~2022M710429.


\begin{thebibliography}{99}
\bibitem{Abbott2016}LIGO Scientific and Virgo Collaborations, Observation of Gravitational Waves from a Binary Black Hole Merger, Phys.Rev.Lett. 116 (2016) 6, 061102 [arXiv: 1602.03837].

\bibitem{Hawking1971}S. W. Hawking, Gravitationally collapsed objects of very low mass, Mon.Not.Roy.Astron.Soc. 152 (1971) 75.

\bibitem{Carr1974}B. J. Carr, S.W. Hawking, Black holes in the early Universe, Mon.Not.Roy.Astron.Soc. 168 (1974) 399-415.

\bibitem{Carr1975}B. J. Carr, The Primordial black hole mass spectrum, Astrophys.J. 201 (1975) 1-19.

\bibitem{CMB2018}Planck collaboration, Planck 2018 results,X. Constraints on inflation, Astron.Astrophys. 641 (2020) A10 [arXiv: 1807.06211].

\bibitem{Sasaki2018}M. Sasaki, T. Suyama, T. Tanaka, S. Yokoyama, Primordial black holes—perspectives in gravitational wave astronomy, Class.Quant.Grav. 35 (2018) 6, 063001 [arXiv: 1801.05235].

\bibitem{Green2021}A. M. Green, B. J. Kavanagh, Primordial Black Holes as a dark matter candidate, J.Phys.G 48 (2021) 4, 043001 [arXiv: 2007.10722].

\bibitem{Cai2020}R.-G. Cai, Z.-K. Guo, J. Liu, L. Liu, Primordial black holes and gravitational waves from parametric amplification of curvature perturbations, JCAP 06 (2020) 013 [arXiv: 1912.10437 ].

\bibitem{Motohashi2020}H. Motohashi, S. Mukohyama, M. Oliosi, Constant Roll and Primordial Black Holes, JCAP 03 (2020) 002 [arXiv: 1910.13235].

\bibitem{Clesse2015}S. Clesse, J. Garc\'ia-Bellido, Massive Primordial Black Holes from Hybrid Inflation as Dark Matter and the seeds of Galaxies, Phys.Rev.D 92 (2015) 2, 023524 [arXiv: 1501.07565]

\bibitem{Cai2019}C. Chen, Y.-F. Cai, Primordial black holes from sound speed resonance in the inflaton-curvaton mixed scenario, JCAP 10 (2019) 068 [arXiv: 1908.03942].

\bibitem{Pi2018}S. Pi, Y.-l. Zhang, Q.-G. Huang, M. Sasaki, Scalaron from $\rm R^2$-gravity as a heavy field, JCAP 05 (2018) 042 [arXiv: 1712.09896].

\bibitem{Fu2019}C.-J. Fu, P.-X. Wu, H.-W. Yu, Primordial Black Holes from Inflation with Nonminimal Derivative Coupling, Phys.Rev.D 100 (2019) 6, 063532 [arXiv: 1907.05042].

\bibitem{Kimura2021}R. Kimura, T. Suyama, M. Yamaguchi, Y.-L. Zhang, Reconstruction of Primordial Power Spectrum of curvature perturbation from the merger rate of Primordial Black Hole Binaries, JCAP 04 (2021) 031 [arXiv: 2102.05280].

\bibitem{Wang2022}X.-P. Wang, Y.-L. Zhang, R. Kimura, M. Yamaguchi, Reconstruction of Power Spectrum of Primordial Curvature Perturbations on small scales from Primordial Black Hole Binaries scenario of LIGO/VIRGO detection, arxiv: 2209.12911.

\bibitem{planck2018}Planck Collaboration, Planck 2018 results. VI. Cosmological parameters, Astron.Astrophys. 641 (2020) A6 [arXiv: 1807.06209].

\bibitem{GWTC3}LIGO Scientific, VIRGO, and KAGRA collaborations, GWTC-3: Compact Binary Coalescences Observed by LIGO and Virgo During the Second Part of the Third Observing Run, arXiv: 2111.03606.

\bibitem{Sasaki2016}M. Sasaki, T. Suyama, T. Tanaka, S. Yokoyama, Primordial Black Hole Scenario for the Gravitational-Wave Event GW150914, Phys.Rev.Lett. 117 (2016) 6, 061101 [arXiv: 1603.08338].

\bibitem{Haimoud2017}Y. Ali-Ha\"{i}moud, E. D. Kovetz, M. Kamionkowski, Merger rate of primordial black-hole binaries, Phys.Rev.D 96 (2017) 12, 123523 [arXiv: 1709.06576].

\bibitem{Raidal2017}M. Raidal, V. Vaskonen, H. Veerm\"{a}e,  Gravitational Waves from Primordial Black Hole Mergers, JCAP 09 (2017) 037 [arXiv:  1707.01480].

\bibitem{Chen2018}Z.-C. Chen, Q.-G. Huang, Merger Rate Distribution of Primordial-Black-Hole Binaries, Astrophys.J. 864 (2018) 1, 61 [arXiv: 1801.10327].

\bibitem{Raidal2018}M. Raidal, C. Spethmann, V. Vaskonen, H. Veerm\"{a}e, Formation and Evolution of Primordial Black Hole Binaries in the Early Universe, JCAP 02 (2019) 018, [arXiv: 1812.01930].

\bibitem{Hutsi2021}G. H\"{u}tsi, M. Raidal, V. Vaskonen, H. Veerm\"{a}e, Two populations of LIGO-Virgo black holes, JCAP 03 (2021) 068 [arXiv: 2012.02786 ].

\bibitem{Luca2021}V. De Luca, G. Franciolini, P. Pani, A. Riotto, Bayesian Evidence for Both Astrophysical and Primordial Black Holes: Mapping the GWTC-2 Catalog to Third-Generation Detectors, JCAP 05 (2021) 003 [arXiv: 2102.03809].

\bibitem{Franciolini2022b}G. Franciolini, I. Musco, P. Pani, A. Urbano, From inflation to black hole mergers and back again: Gravitational-wave data-driven constraints on inflationary scenarios with a first-principle model of primordial black holes across the QCD epoch, arXiv: 2209.05959.

\bibitem{Carr2017}B. Carr, M. Raidal, T. Tenkanen, V. Vaskonen, Primordial black hole constraints for extended mass functions, Phys.Rev.D 96 (2017) 2, 023514 [arXiv: 1705.05567].

\bibitem{Bellomo2018}N. Bellomo, J. L. Bernal, A. Raccanelli, Licia Verde, Primordial Black Holes as Dark Matter: Converting Constraints from Monochromatic to Extended Mass Distributions, JCAP 01 (2018) 004 [arXiv: 1709.07467].

\bibitem{Green2016}A. M. Green, Microlensing and dynamical constraints on primordial black hole dark matter with an extended mass function, Phys.Rev.D 94 (2016) 6, 063530 [arXiv: 1609.01143].

\bibitem{Kannike2017}K. Kannike, L. Marzola, M. Raidal, H. Veerm\"{a}e, Single Field Double Inflation and Primordial Black Holes, JCAP 09 (2017) 020 [arXiv: 1705.06225].

\bibitem{Talbot2018}C. Talbot, E. Thrane, Measuring the binary black hole mass spectrum with an astrophysically motivated parameterization, Astrophys.J. 856 (2018) 2, 173 [arXiv: 1801.02699].

\bibitem{LVK2022}LIGO Scientific, VIRGO, and KAGRA collaborations, The population of merging compact binaries inferred using gravitational waves through GWTC-3, arXiv: 2111.03634.

\bibitem{Madau2014}P. Madau, M. Dickinson, Cosmic Star-Formation History, Ann.Rev.Astron.Astrophys. 52 (2014) 415-486 [arXiv: 1403.0007].

\bibitem{Fishbach2018}M. Fishbach, D. E. Holz,  W. M. Farr, Does the Black Hole Merger Rate Evolve with Redshift? Astrophys.J.Lett. 863 (2018) 2, L41 [arXiv: 1805.10270].

\bibitem{Mandel2019}I. Mandel, W. M. Farr, J. R. Gair, Extracting distribution parameters from multiple uncertain observations with selection biases, Mon.Not.Roy.Astron.Soc. 486 (2019) 1, 1086 [arXiv: 1809.02063].

\bibitem{Chen2020}Z.-C. Chen,  Q.-G. Huang, Distinguishing Primordial Black Holes from Astrophysical Black Holes by Einstein Telescope and Cosmic Explorer, JCAP 08 (2020) 039 [arXiv: 1904.02396].

\bibitem{Chen2019}Z.-C. Chen, F. Huang, Q.-G. Huang, Stochastic Gravitational-wave Background from Binary Black Holes and Binary Neutron Stars and Implications for LISA, Astrophys.J. 871 (2019) 1, 97 [arXiv: 1809.10360].

\bibitem{Luca2020}V. De Luca, G. Franciolini, P. Pani, A. Riotto, Primordial Black Holes Confront LIGO/Virgo data: Current situation, JCAP 06 (2020) 044 [arXiv: 2005.05641].

\bibitem{Wong2021}K. W. K. Wong, G. Franciolini, V. De Luca, et al., Constraining the primordial black hole scenario with Bayesian inference and machine learning: the GWTC-2 gravitational wave catalog, Phys.Rev.D 103 (2021) 2, 023026 [arXiv: 2011.01865].

\bibitem{Wu2021}Y. Wu, Merger history of primordial black-hole binaries, Phys.Rev.D 101 (2020) 8, 083008 [arXiv: 2001.03833].

\bibitem{Chen2022}Z.-C. Chen,  S.-S. Du, Q.-G. Huang,  Z.-Q. You, Constraints on Primordial-black-hole Population and Cosmic Expansion History from GWTC-3, arXiv:2205.11278.

\bibitem{Liu:2022iuf}L.~Liu, Z.~Q.~You, Y.~Wu and Z.~C.~Chen,
Constrain the Merger History of Primordial-Black-Hole Binaries from GWTC-3, arXiv:2210.16094.

\bibitem{Franciolini2022}G. Franciolini, V. Baibhav, V. De Luca, K. K. Y. Ng, K. W. K. Wong, E. Berti, P. Pani, A. Riotto and S. Vitale, Searching for a subpopulation of primordial black holes in LIGO-Virgo gravitational-wave data, Phys. Rev. D 105, 8, 083526 (2022) [arXiv:2105.03349].

\bibitem{Ng2022}K. K.Y. Ng, G. Franciolini, E. Berti, et al., Constraining High-redshift Stellar-mass Primordial Black Holes with Next-generation Ground-based Gravitational-wave Detectors, Astrophys.J.Lett. 933 (2022) 2, L41 [arXiv: 2204.11864].

\bibitem{LIGOo3}LIGO Scientific Collaboration, Virgo Collaboration, and KAGRA Collaboration, GWTC-3: Compact Binary Coalescences Observed by LIGO and Virgo During the Second Part of the Third Observing Run — O3 search sensitivity estimates, (2021), https://doi.org/10.5281/zenodo.5546676.

\bibitem{icarogw}S. Mastrogiovanni, K. Leyde, C. Karathanasis, E. Chassande-Mottin, D. A. Steer, J. Gair, A. Ghosh, R. Gray, S. Mukherjee, and S. Rinaldi,  On the importance of source population models for gravitational wave cosmology,  Phys. Rev. D 104, 062009 (2021), [arXiv:2103.14663].

\bibitem{bilby}I. M. Romero-Shaw et al., Bayesian inference for compact binary coalescences with bilby: validation and application to the first LIGO–Virgo gravitational-wave transient catalogue, Mon. Not. Roy. Astron. Soc. 499, 3295–3319 (2020) [arXiv:2006.00714].

\bibitem{Jeffreys}H. Jeffreys, The theory of probability (3rd ed.), Oxford, England (1998).

\bibitem{Schwarz1978}G. Schwarz, Estimating the Dimension of a Model, Annals Statist. 6 (1978) 461.

\bibitem{Akaike1974}H. Akaike, A new look at the statistical model identification, IEEE Trans. Automat. Contr. 19 716 (1974). 

\bibitem{Musco2013}I. Musco, J. C. Miller, Primordial black hole formation in the early universe: critical behaviour and self-similarity, Class.Quant.Grav. 30 (2013) 145009 [arXiv: 1201.2379].

\bibitem{Harada2014}T. Harada, C.-M. Yoo, K. Kohri, Threshold of primordial black hole formation, Phys.Rev.D 89 (2014) 2, 029903 [arXiv: 1309.4201].

\bibitem{Yoo2018}C.-M. Yoo, T. Harada, J. Garriga, K. Kohri, Primordial black hole abundance from random Gaussian curvature perturbations and a local density threshold, PTEP 2018 (2018) 12, 123E01 [arXiv: 1805.03946].  
\bibitem{Gow2020}A. D. Gow, C. T. Byrnes, P. S. Cole, S. Young, The power spectrum on small scales: Robust constraints and comparing PBH methodologies, JCAP 02 (2021) 002 [arXivt: 2008.03289].

\bibitem{Young2013}S. Young, C. T. Byrnes, Primordial black holes in non-Gaussian regimes, JCAP 08 (2013) 052 [arXiv: 1307.4995].

\bibitem{Pattison2017}C. Pattison, V. Vennin, H. Assadullahi, D. Wands, Quantum diffusion during inflation and primordial black holes, JCAP 10 (2017) 046 [arXiv: 1707.00537].

\bibitem{Franciolini2018}G. Franciolini, A. Kehagias, S. Matarrese, A. Riotto, Primordial Black Holes from Inflation and non-Gaussianity, JCAP 03 (2018) 016 [arXiv: 1801.09415].

\bibitem{Atal2019}V. Atal, C. Germani, The role of non-gaussianities in Primordial Black Hole formation, Phys.Dark Univ. 24 (2019) 100275 [arXiv: 1811.07857].

\bibitem{Luca2022}V. De Luca, A. Riotto, A note on the abundance of primordial black holes: Use and misuse of the metric curvature perturbation, Phys.Lett.B 828 (2022) 137035 [arXiv: 2201.09008].

\bibitem{Kalaja2019}A. Kalaja,  N. Bellomo, N. Bartolo, D. Bertacca, S.Matarrese, A. Raccanelli, L. Verde, From Primordial Black Holes Abundance to Primordial Curvature Power Spectrum (and back), JCAP 10 (2019) 031[arXiv: 1908.03596].

\bibitem{Luca2019}V. De Luca, G. Franciolini, A. Kehagias, M. Peloso, A. Riotto, C. \"Unal, The Ineludible non-Gaussianity of the Primordial Black Hole Abundance, JCAP 07 (2019) 048 [arXiv: 1904.00970].

\bibitem{Young2019}S. Young, I. Musco, C. T. Byrnes, Primordial black hole formation and abundance: contribution from the non-linear relation between the density and curvature perturbation, JCAP 11 (2019) 012 [arXiv: 1904.00984].

\end{thebibliography}
\end{document}